\documentclass[acmlarge,screen]{acmart}
\AtBeginDocument{%
  }

\setcopyright{acmlicensed}
\copyrightyear{2018}
\acmYear{2018}
\acmDOI{XXXXXXX.XXXXXXX}
\acmConference[Conference acronym 'XX]{Make sure to enter the correct
  conference title from your rights confirmation email}{June 03--05,
  2018}{Woodstock, NY}
\acmISBN{978-1-4503-XXXX-X/2018/06}

\begin{document}

\title{Characterizing Unintended Consequences in Human-GUI Agent Collaboration for Web Browsing}

\author{Shuning Zhang}
\orcid{0000-0002-4145-117X}
\email{zsn23@mails.tsinghua.edu.cn}
\affiliation{%
  \institution{Tsinghua University}
  \city{Beijing}
  \country{China}
}

\author{Jingruo Chen}
\orcid{0009-0007-1606-5780}
\email{jc3564@cornell.edu}
\affiliation{%
  \institution{Information Science, Cornell University}
  \city{Ithaca}
  \state{New York}
  \country{USA}
}

\author{Zhiqi Gao}
\orcid{0009-0008-4898-2521}
\email{gaozhiqi@mail.nankai.edu.cn}
\affiliation{%
  \institution{Nankai University}
  \city{Tianjin}
  \country{China}
}

\author{Jiajing Gao}
\affiliation{%
 \institution{Institute of Future Human Habitat, Tsinghua University}
 \city{Shenzhen}
 \country{China}}

\author{Xin Yi}
\orcid{0000-0001-8041-7962}
\authornote{Corresponding author.}
\email{yixin@tsinghua.edu.cn}
\affiliation{
    \institution{Tsinghua University}
    \city{Beijing}
    \country{China}
}
\affiliation{
    \institution{Zhongguancun Laboratory}
    \city{Beijing}
    \country{China}
}

\author{Hewu Li}
\orcid{0000-0002-6331-6542}
\email{lihewu@cernet.edu.cn}
\affiliation{
    \institution{Tsinghua University}
    \city{Beijing}
    \country{China}
}
\affiliation{
    \institution{Zhongguancun Laboratory}
    \city{Beijing}
    \country{China}
}

\renewcommand{\shortauthors}{Trovato et al.}

\begin{abstract}
  % The rising use of Large Language Model (LLM)-based Graphical User Interface (GUI) agents, especially in browsing scenarios, presents complex unintended consequences (UCs). This paper characterizes these UCs from four perspectives: phenomena, influences, reasons, and mitigation. We combined social media analysis (N=221 posts) and semi-structured interviews (N=14) to triangulate the results. Our analysis was embedded within the context of agents' working process, the vulnerability exposure of GUI agents, human emotional responses, various stakeholders, and the broad social-technical backgrounds. The phenomena identified include privacy and consent issues, unanticipated actions, amplifying misinformation, and failure to complete operations. These lead to influences such as security and privacy concerns, financial and social loss, user experience frustration, and denial of service. User-initiated mitigation strategies involve technical improvements and API integration, manual correction, error detection and prevention, and environment restrictions. Our results provide four insightful implications for designing future LLM-based GUI agents that are more robust, user-centric and transparent, inspiring a balance between automation and human oversight.
  The proliferation of Large Language Model (LLM)-based Graphical User Interface (GUI) agents in web browsing scenarios present complex unintended consequences (UCs). This paper characterizes three UCs from three perspectives: phenomena, influences and mitigation, drawing on social media analysis (N=221 posts) and semi-structured interviews (N=14). Key phenomenon for UCs include agents' deficiencies in comprehending instructions and planning tasks, challenges in executing accurate GUI interactions and adapting to dynamic interfaces, the generation of unreliable or misaligned outputs, and shortcomings in error handling and feedback processing. These phenomena manifest as influences from unanticipated actions and user frustration, to privacy violations and security vulnerabilities, and further to eroded trust and wider ethical concerns. Our analysis also identifies user-initiated mitigation, such as technical adjustments and manual oversight, and provides implications for designing future LLM-based GUI agents that are robust, user-centric, and transparent, fostering a crucial balance between automation and human oversight.
\end{abstract}



%   \includegraphics[width=0.8\textwidth]{Figure/gui-agent.png}
%   \caption{The .}

\received{20 February 2007}
\received[revised]{12 March 2009}
\received[accepted]{5 June 2009}

\maketitle

\section{Introduction}

With the increasing prevalence of Large Language Models (LLMs), their application is expanding beyond simple interfaces towards enabling new forms of computer-supported cooperative work (CSCW). A promising direction involves LLM-based agents capable of interacting with Graphical User Interfaces (GUIs)~\cite{gao2024assistgui}, alongside multi-agent systems for complex tasks~\cite{qian2024chatdev} and social simulations~\cite{park2023generative}. Among these, GUI agents, exemplified by commercial products such as Computer Use\footnote{\url{https://www.anthropic.com/news/3-5-models-and-computer-use}} and Operator\footnote{\url{https://openai.com/index/introducing-operator/}} and open-source projects like UI-TARS\footnote{\url{https://github.com/bytedance/UI-TARS}} and Open Interface\footnote{\url{https://github.com/AmberSahdev/Open-Interface}}, are rapidly gaining traction.

\begin{figure*}
    \includegraphics[width=0.8\textwidth]{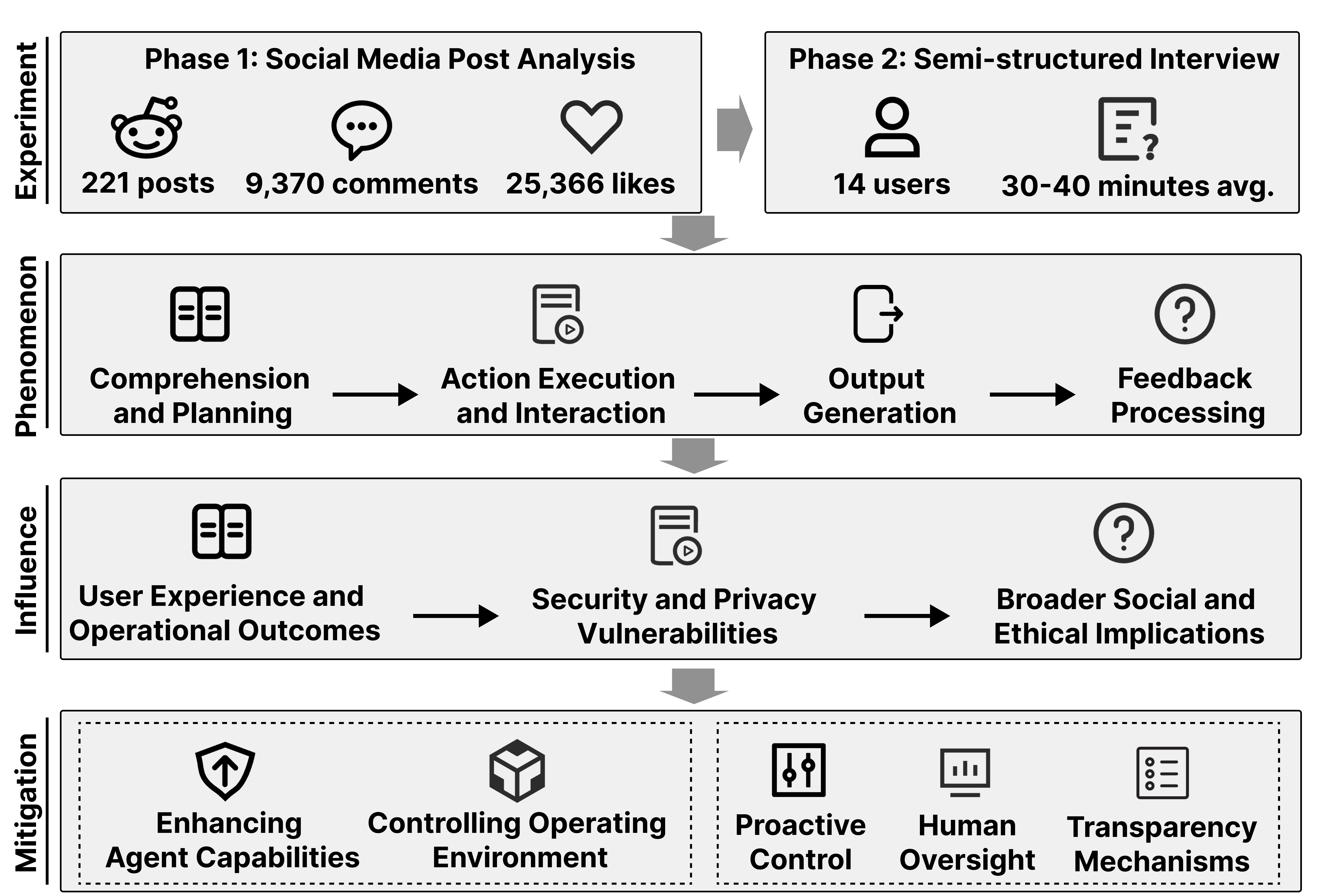}
    \caption{The framework of this paper. Through the two-phase studies, we analyzed the UCs of GUI agents from its phenomenon, influence and mitigation.}
    \label{fig:frame_gui_agent}
\end{figure*}

These GUI agents act as proxies or digital collaborators, automating operations within the user's shared digital workspace. They can perform tasks such as sending tweets\footnote{See examples from UI-TARS repository: \url{https://github.com/bytedance/UI-TARS}.}, retrieving weather information\footnote{See examples from UI-TARS repository: \url{https://github.com/bytedance/UI-TARS}.}, or making dinner plans\footnote{See examples from Open Interface repository: \url{https://github.com/AmberSahdev/Open-Interface}.}. \textbf{Web browsing, involving information searching, purchasing, messaging, and interacting with web applications,} represents a particularly significant domain for this human-agent cooperation. While theoretical and technical discussions highlight the burgeoning capabilities of these agents on benchmarks or curated task environments~\cite{nguyen2024gui,zhao2024gui}, their practical deployment and failures remain less understood, especially considering that the integration of these agents into human workflows may results in unintended consequences (UCs), such as purchasing incorrect items or quantities, or misdirecting messages on web platforms. Within the scope of this investigation, UCs are \textbf{conceptualized as unanticipated, negative outcomes in the interactive processes of these GUI agents.}

This conceptualization of UCs specific to GUI agents aligns with, yet extends prior research on  technology-induced harms. For instance, prior work detailed issues such as snooping and misuse on smartphones~\cite{marques2016snooping,marques2014measuring,muslukhov2013know,arikewuyo2021influence} and personal computers~\cite{frampton2021monitoring,hawk2016snooping}, often involving monitoring by social insiders. Vulnerabilities and misuse related to IoT devices have also been explored~\cite{oksman2010mobile}. While this body of work provides valuable insights into technology misuse in various contexts, the specific examination of GUI agents operating in a dynamic collaborative environment remains unexplored. 

The imperative to address this research gap becomes particularly evident considering the multifaceted and severe nature of UCs impacting human-GUI agent collaboration. These UCs seriously impair operational integrity, as agents misinterpret objectives to derail research or execute tasks flawedly; they strain resource management, by incurring financial losses from erroneous purchases or disrupting workflows through excessive confirmations; and they compromise user security, via critical privacy breaches from mishandled data or the spread of misinformation. Far from trivial errors, these incidents signify deep-seated challenges in human-agent web collaboration, demanding substantive solutions. To systematically dissect these complex issues, a structured approach is indispensable. This research, therefore, adopts an analytical framework centered on sequentially investigating the phenomenon of these UCs, their ensuing influence, and potential mitigation strategies. Such a structured investigation allows for the characterization of observable issues, an understanding of their multi-level impacts, and the exploration of solutions. Consequently, we aim to address:
% For instance, an agent designed to assist with web research might derail the process by navigating to irrelevant sites. Similarly, an agent intended to facilitate online purchases could make costly errors. Users might also experience significant workflow disruptions when agents demand excessive confirmations for routine web actions, or face privacy breaches and the spread of misinformation due to agents improperly handling web-sourced data. Such incidents are not merely trivial software errors, but represent deep-seated challenges in human-agent collaboration within dynamic web environments that necessitate more than superficial solutions. Understanding these varied consequences is crucial for designing systems that support the complex dynamics of human-agent cooperative work in web browsing environments. Therefore, we aim to characterize these distinct issues by addressing:

\textbf{[Phenomenon] RQ1. What UCs do users face when using GUI agents for web browsing?}

\textbf{[Influence] RQ2. What are the influences of UCs from the user's perspective during GUI agents' web browsing?}

\textbf{[Mitigation] RQ3. How do users attempt to mitigate the consequences and what's users' expectations?}

To address these inter-connected research questions, we conducted a series of studies, involving social media analyses (N=221 posts) and semi-structured interviews (N=14), similar to previous practices~\cite{lukoff2021design,alberts2024computers}. 
Towards RQ1, we characterized the phenomena of UCs by analyzing UCs across the agent's operational lifecycle. Our findings reveal critical breakdowns occurred during the \textit{Input Stage} (e.g., failures in comprehension, planning and instruction adherence), the \textit{Action Stage} (e.g., erroneous GUI manipulations, inability to adapt to dynamic UIs, element localization errors), the \textit{Output Stage} (e.g., generation of inaccurate or semantically unsatisfactory results), and the \textit{Feedback Stage} (e.g., deficient error handling, slow system response, prohibitive costs). 

Towards RQ2, building upon the identified failures, we examined the influences of these UCs across three progressive levels. On the basic level, UCs have direct impacts on task execution and user experience, including task failures, operational inefficiencies, increased user workload and frustration. Furthermore, UCs could lead to security threats and privacy violations, encompassing risks like unauthorized data access, exposure, vulnerability to malicious exploitation, surveillance and social engineering. These culminate in broader consequences for trust, adoption and society, characterized by eroded user trust, hindered technology adoption and wider socio-ethical concerns.

Finally, towards RQ3, we examined how users attempt to mitigate these consequences, and their expectations for future agents. Users propose system-oriented improvements aimed at enhancing agent robustness (e.g., better capabilities, intrinsic safety limits) and formalizing operational controls (e.g., standardized restrictions, robust permissions). Simultaneously, users engage in various user-oriented mitigation practices, including careful management of agent input and scope, direct oversight and intervention during execution, and configuring the operational environment (e.g., using isolated systems). While these practices demonstrate user adaptation, they often underscore current agent limitations and involve considerable user effort. We also gathered user expectations emphasizing the need for enhanced reliability, controllability, transparency, personalization, security, and the potential benefits of localized processing, leaving several design implications. To sum up, the contributions of this paper are threefold:

\begin{itemize}
    \item We contributed an initial characterization of UCs' phenomena around GUI agents in web browsing.
    \item Our analysis of the multifaceted influences of these UCs reveals their significant, yet previously underexplored, impact on users and system interaction.
    \item We synthesized user-initiated mitigation strategies and proposed design implications, offering guidance for the responsible advancements of GUI agents.
\end{itemize}

\section{Related Work}

This section synthesizes three research streams vital to the CSCW community: the study of UCs from emerging socio-technical systems~\cite{flanagan2014making,lalone2014values,fiesler2023internet}, the advancements and limitations of LLM agents~\cite{chen2024exploring}, and the security, privacy and safety risks associated with their deployment, especially in collaborative interactions~\cite{vitak2021designing,jung2023toward}.

\subsection{Unintended Consequences of Emerging Techniques}

The CSCW community pioneered the study of UCs stemming from design actions and socio-technical systems~\cite{flanagan2014making}. Early work examined how overlooked design elements, or `values levers', could facilitate negative societal impacts, such as the white nationalist appropriation of content~\cite{lalone2014values}. Subsequent CSCW research broaden this scope, investigating UC challenges across various domains, including the lifecycle of online research~\cite{fiesler2023internet}. More recently, Chen et al.~\cite{chen2024exploring} explored user attribution of responsibility for UCs with AI systems. While foundational, this body of work does not specifically address the nuanced failures modes of contemporary GUI agents in collaborative web environments, which is our focus.

Research across HCI, usable security and software engineering documents divides UCs from technology design and misuse. Studies highlight issues such as widespread snooping~\cite{marques2016snooping,marques2014measuring,muslukhov2013know,arikewuyo2021influence}, unauthorized access~\cite{marques2016snooping}, data manipulation~\cite{karlson2009can} and threats involving smartphones and personal/shared computers~\cite{frampton2021monitoring,hawk2016snooping,carter2002misuse}. Multi-user designs in emerging technologies like IoT devices can also expand misuse surfaces, particularly in settings like smart homes~\cite{oksman2010mobile,moh2023characterizing}. Furthermore, investigations reveal that security and privacy (S\&P) software can generate significant negative UCs~\cite{ramulu2024security}. These documented outcomes including creating access barriers~\cite{fassl2022comparing,namara2020emotional}, engaging in misleading marketing~\cite{fassl2022comparing,akgul2022investigating}, presenting accessibility hurdles~\cite{renaud2020dyslexia,huaman2022they}, exhibiting poor usability, facilitating abuse like stalking~\cite{eterovic2017stalking,gallardo2022detecting}, and disproportionately harming vulnerable populations~\cite{mcdonald2020privacy}. Although prior work reveals diverse UCs, it overlooks those specific to proactive GUI agents in collaborative web environments, which is our research focus.

Acknowledging these risks~\cite{dorin2021ethical}, researchers have proposed mitigation strategies, including privacy frameworks~\cite{wright2012privacy} and design methodologies~\cite{elsayed2023responsible}. However, UCs persist, partly due to implementation challenges such as developers' limited ethical training and domain expertise~\cite{hadar2018privacy,fulton2023vulnerability}. Therefore, focusing on characterizing the problem, we contribute an initial categorization of GUI agent UCs from the user perspective, aiming to inform subsequent countermeasure development.

\subsection{Task Automation and GUI Agents}

Traditional GUI automation primarily employed rule-based frameworks such as Selenium\footnote{\url{https://www.selenium.dev/documentation/}}, Robot Framework\footnote{\url{https://robotframework.org/robotframework/}} and AutoIt\footnote{\url{https://www.autoitscript.com/site/autoit/documentation-localization/}}~\cite{chen2025obvious}. These tools automate predefined interaction sequences (e.g., clicks, text input) but exhibit limited adaptability in dynamic environments and necessitate substantial manual configuration.

The advent of LLMs catalyzed a paradigm shift, fostering sophisticated GUI agents engineered for real-time GUI component interpretation and dynamic adaptation to interface modifications (e.g., layout changes, content updates). These agents leverage LLMs to comprehend visual interfaces and autonomously execute user instructions by emulating human-like interaction patterns such as clicking and typing~\cite{wang2024gui}. Such advancements spurred research into enhanced GUI automation, exemplified by LLM-powered systems for flexible testing~\cite{li2024test}, frameworks incorporating symbolic reasoning to refine GUI interactions~\cite{judson2024soid}, specialized mobile application automation~\cite{zhang2025appagent}, and the direct translation of natural language instructions into executable GUI actions~\cite{huangprompt2task}.

% The advent of LLMs catalyzed a paradigm shift, fostering the development of sophisticated GUI agents. These agents are engineered to interpret GUI components in real-time, thereby enabling dynamic adaptation to interface modifications like layout changes or content updates. For instance, Li et al.~\cite{li2024test} introduced an LLM-powered system enhancing GUI testing flexibility. Concurrently, Judson et al.~\cite{judson2024soid} investigated symbolic reasoning within automated decision-making frameworks to refine GUI interactions, particularly for scenarios requiring robust accountability.

% At their core, GUI agents leverage LLMs to comprehend visual interfaces and autonomously execute user instructions by emulating human-like interaction patterns, such as clicking and typing~\cite{wang2024gui}. This fundamental capability has spurred the creation of specialized task automation frameworks. Notable examples include systems designed for mobile application interaction~\cite{zhang2025appagent} and those that translate natural language instructions directly into executable GUI actions~\cite{huangprompt2task}. 

Despite advancements, the practical deployment of GUI agents, including commercially explored variants (e.g., OpenAI~\cite{openai_operator}, Anthropic~\cite{anthropic_computer_use}), reveals persistent operational deficiencies. These specific shortcomings of GUI agents in real-world settings remain insufficiently examined, distinct from research addressing broader AI agent limitations~\cite{hu2024dawn} or detailing isolated implementations. Consequently, a focused investigation into these unique operational challenges is warranted.

\subsection{Security, Privacy and Safety Risks in Agent Usage}

Research on agent-related risks spans various domains. Within CSCW, studies have investigated privacy-enhancing design principles~\cite{vitak2021designing} and the limitations of AI value assessment methodologies~\cite{jung2023toward}. Broader AI research has contributed by categorizing security threats~\cite{deng2025ai}, outlining design challenges based on expert perspectives~\cite{li2024personal} and designing specific privacy protection systems~\cite{zhang2024ghost,zhang2024adanonymizer}. Concurrently, technical efforts have focused on developing specific mitigation strategies, such as constrained environments like AirGapAgent~\cite{bagdasarian2024airgapagent}, and frameworks for testing tool integration, exemplified by ToolEmu~\cite{ruan2023identifying}. However, these work did not address the unique risks when LLMs drive GUI agents for interactive tasks. 

For instance, while foundational studies like Deng et al.~\cite{deng2025ai} offer a general threat taxonomy for AI systems (encompassing perception, reasoning, action, and memory components) and Li et al.~\cite{li2024personal} detail LLM security and privacy concerns such as confidentiality, integrity and reliability, their analyses do not delineate the distinct vulnerabilities or safety implications arising specifically from GUI agent capabilities and interactions. Other specialized research, including investigations into risks inherent to scientific LLM agents~\cite{tang2024prioritizing} or surveys focused on AI agent memory mechanisms~\cite{zhang2024survey}, also diverges from the specific context of LLM-driven GUI control. Consequently, a comprehensive understanding of the security, privacy and safety challenges unique to the operational use of GUI agents remain under-developed.

\section{Methods}

Our studies employed an exploratory mixed qualitative methodology~\cite{lukoff2021design,o2021mixing,tran2019modeling}, also characterized as an intra-paradigm approach~\cite{oaa2015advanced}, to investigate user experiences with GUI agents during web browsing. The research was conducted in two primary stages: a social media analysis (N=221 posts) followed by a semi-structured in-depth interview (N=14). The latter phase used different participants to reach diverse and deep results. All interviews and analysis took place online.

To synthesize our findings, we integrated qualitative data for triangulation. We intentionally designed this multi-method strategy to mitigate the limitations inherent in any single method and to foster a nuanced understanding of participants' experiences and preferences. We iteratively performed thematic analysis of the integrated dataset, drawing upon Braun and Clarke's~\cite{braun2019reflecting,braun2024thematic} reflexive methodology and O'Reilly et al's~\cite{o2021mixing} integrated analytic approach. Through this exploratory approach, we aimed to identify and characterize UCs, their influences, users' mitigation and expectations concerning GUI agents. These studies received the ethical approval from our university's Institutional Review Board (IRB).

% Our study had two phases: a social media analysis with 221 posts, followed by a semi-structured in-depth interview (N=14). The latter phase used different participants to reach more diverse results and deepen the results from an orthogonal perspective. All interviews and workshops took place online.

% Using a mixed qualitative method~\cite{lukoff2021design,o2021mixing,tran2019modeling} (or intra-paradigm~\cite{oaa2015advanced}) design, we integrated qualitative data from the four phrases to triangulate our findings. This combination of methods was designed to counteract the limitations of each phase, and help us to uncover nuances in our participants' experiences and preferences. Data was thematically analyzed in an iterative, integrated way using a combination of Braun and Clarke's~\cite{braun2019reflecting,braun2024thematic} reflexive approach and O'Reilly et al.'s~\cite{o2021mixing} integrated analytic approach, described below. Our aim was exploratory: to start identifying the UCs of GUI agents during web browsing. We specifically focused on the phenomena, concerns, causes, expectations, etc. This study was approved by our university's Institutional Review Board (IRB). Publicly shareable data, coding, and study materials would be available upon acceptance.

\subsection{Phase 1: Social Media Analysis}

We first carried out a social media analysis on Reddit~\footnote{\url{https://www.reddit.com/}} to analyze the general opinions of users towards GUI agents. We used keywords \textit{Operator}, \textit{Computer Use} (and \textit{computer-use}), \textit{UI-TARS}, \textit{mistake}, \textit{fault}, \textit{ai agent}, and their combinations to search through Reddit, as we found that searching only in subreddits like \textit{r/Operator} or \textit{r/aiagents} are not representative and comprehensive enough. We went through all posts (1,850 in total) and manually screened these posts' titles and content to determine whether they are correlated with GUI agents. We also deleted a few whose content was not related to UCs of GUI agents (e.g., introducing the functions of GUI agents). This screening process excluded 1,629 posts and left 221 posts (25,366 likes) and 9,370 comments (with 42 comments on average for each post, SD=65.3, see Appendix~\ref{appendix:social} for detailed distribution across sub-reddits). As most comments are discussed around a specific topic, and for ease of annotation, we reported the results at the level of posts.

\subsection{Phase 2: Semi-structured Interview}

As the social media analysis may be biased from exposure bias~\cite{kopec1990bias} and may incentivize concise responses, this phase enabled participants to elaborate on the responses with UCs. We recruited participants through distributing posters on online social media including Reddit, the Redbook\footnote{\url{https://www.xiaohongshu.com/explore}}, and WeChat\footnote{\url{https://www.wechat.com/}}. Our recruitment lasted two weeks. After screening, we selected 14 participants with GUI agents' experience (see Table~\ref{tab:dataoccupations}). They were diverse in occupations and usage experiences, comprising 12 males and 2 females. 

To enhance data validity, participants were instructed pre-interview to briefly re-familiarize themselves with their previously used GUI agent(s), a preparatory step intended to refresh their recall of UCs. We conducted 1:1 semi-structured interviews (30 - 40 minutes) with each participant over Tencent meeting\footnote{\url{https://meeting.tencent.com/}}. We designed the interview protocol according to the social media analysis, asking participants whether it happens in their daily lives, and asked them to elaborate if it happened. The interview script could be found in Appendix~\ref{appendix: interview protocol}.

\begin{table}[!htbp]
\centering
\caption{Participants' demographics.}
\label{tab:dataoccupations}
\begin{tabular}{lll}
\toprule
\textbf{Participant} & \textbf{Experience} & \textbf{Occupation} \\
\midrule
I1 & Manus & Master Student\\
\hline
I2 & GLM PC, GLM Agent & Tech Product Manager\\
\hline
I3 & Operator, Computer Use & Tech Entrepreneur\\
\hline
I4 & Bot.new, cursor, Computer Use, Browser Use, Omniparser & Software Engineer\\
\hline
I5 & Not disclosed & Investment Manager\\
\hline
I6 & GLM Agent & Chief Information Officer\\
\hline
I7 & Auto GPT, Operator & Undergraduate Student\\
\hline
I8 & Operator & Director of Technology\\
\hline
I9 & GLM Agent & Software Engineer\\
\hline
I10 & GLM PC & Master Student\\
\hline
I11 & GLM Agent, Manus & PhD Student\\
\hline
I12 & Not disclosed & Machine Learning Engineer\\
\hline
I13 & Open Interpreter, GLM Agent, UI-TARS, Manus & Faculty\\
\hline
I14 & OpenAI Operator, Browser Use, Open WebUI & Consultant\\
\bottomrule
\end{tabular}
\end{table}

\section{RQ1: Observed Behaviors and Operational Failures of GUI Agents}\label{sec:4}

\begin{figure}[!htbp] 
    \centering
    \includegraphics[width=0.7\textwidth]{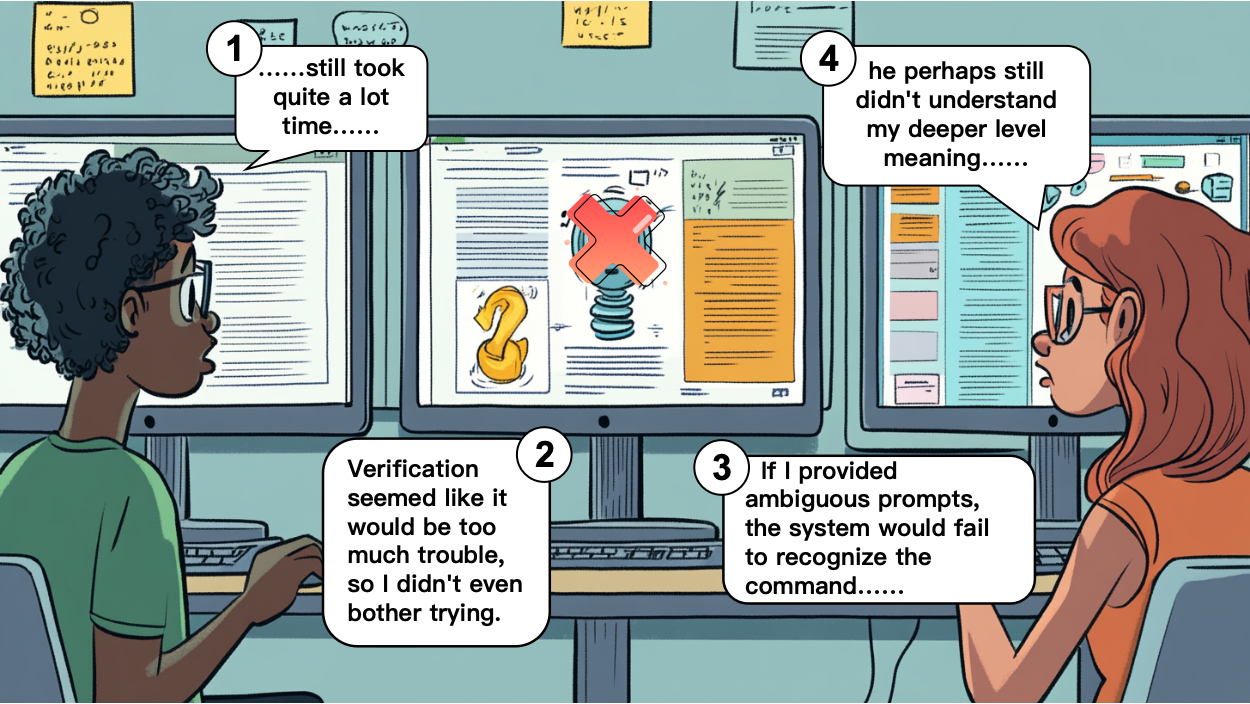}
    \caption{The illustration of the failures users encountered in the (1) input (comprehension/planning), (2) action (execution/interaction), (3) output (generation), and (4) feedback (adjustment/viability) stages. The picture is edited using Midjourney.}
    \label{fig:rq1}
\end{figure}

To address RQ1, we first identify the tasks participants performed using GUI agents. Subsequently, we present a detailed taxonomy of observed operational failures, categorizing them according to the four primary stages of the agent interaction lifecycle: Input (comprehension/planning), Action (execution/interaction), Output (generation), and Feedback (adjustment/viability) (illustrated in Figure~\ref{fig:rq1}).

\subsection{Tasks Participants Used GUI Agents to Complete}\label{sec:4_1}

Understanding the range of tasks users attempted provides context for identifying where and how these failures occurred. Participant tasks spanned six primary areas, encompassing \textbf{navigation and travel planning} (I5, I7), such as route planning and ticket booking (e.g., voice commands yielding complete trip plans (I5)); and \textbf{food and beverage ordering} (I5, I6, I9, I11, I12), where agents expedited takeout based on specified preferences (I5). Users also engaged in \textbf{e-commerce, product research, and shopping} (I11, I12, I13), which included product research and online purchasing (e.g., an agent navigating JD.com for a drone (I3)). Further tasks involved \textbf{information retrieval and web automation} (I7, I10, I14), such as website searches and automated data gathering, with agents independently navigating sites for topics (I4); \textbf{social media management and communication} (I5, I11, I12, I14), like posting and scheduling messages (e.g., timed WeChat messages (I11)); and finally, \textbf{local services and daily task automation} (I6, I12), for instance, using apps for amenities or activating shared bicycles.

\subsection{Failures in Comprehension and Planning (Input Stage)}\label{sec:4_2}

Agent fail during the input stage, hindering task initiation and comprehension through four primary issues, including: (1) setup and configuration complexities, which obstruct initial user access and engagement, (2) inadequate task decomposition, preventing the formulation of a viable action plan, (3) instruction misinterpretations, causing fundamental misunderstanding of user intent, and (4) knowledge deficiencies, impairing the context required to plan.

\subsubsection{Complex Setup and Configuration}\label{sec:4_2_1}
% \subsubsection{Cumbersome Setup Friction}

Significant user friction arises from cumbersome setup procedures and subsequent operational demands, manifesting in several key areas. Users reported substantial initial technical investment, such as \textit{``for this agent I even had to set up a Windows virtual machine'' (I4)} merely to operate the agent. The deployment and configuration stages were often also protracted and complex, comprehensively deploying all aspects reportedly \textit{``still took quite a lot time,''} and even then, ongoing use was found to be \textit{``operationally still relatively troublesome'' (I6)}. Even enabling advanced functionalities necessitated extensive and burdensome upfront preparation, a prerequisite described by users as very troublesome (I8). Beyond these hurdles, specific operational characteristics further contributed to user frustration. Agents' tendency to monopolize the user interface during processing, where \textit{``it occupies my UI interface,''} could exhaust user patience \textit{``even if it processes efficiently''} (I2). 

\subsubsection{Failures in Task Decomposition}\label{sec:4_2_2}

Agents frequently struggle with deconstructing complex instructions into manageable steps and adhering to intended operational sequences. For example, an agent given a complex task like route planning failed to complete the task. According to I2, it \textit{``initiated the task by launching the Maps application but halted immediately afterward, unable to proceed with subsequent necessary actions.''} This inability to follow a logical progression is also seen when agents deviate from specified workflows. I3 reported that when tasked with accessing social media platform, the agent inconsistently navigated to either the correct official website or irrelevant auxiliary portals like its open platforms or service platform, failing to correctly execute even seemingly straightforward sub-tasks. 

Agents also struggled with complex UI manipulations. I4 described scenarios involving multiple clicks, selections, data entry (such as ``adding this''), and pagination (``needs to turn pages''), which posed severe challenges for agents, often leading to incomplete or incorrect task execution. Moreover, agents exhibited more fundamental execution errors. I10 recounted instances where agents failed to respond to double-clicks, became stuck midway through a process, or skipped critical steps during automation. I11 similarly observed that agents could become completely unresponsive, \textit{``just stopping there and becoming unresponsive.''} The issue of looping behavior was another frequent concern. I13 described how an agent would become trapped in a repetitive cycle, \textit{``getting stuck in some kind of loop, continuously clicking into something and then backing out for a very long time.''}

\subsubsection{Instruction Misinterpretation}\label{sec:4_2_3}
Agents frequently misinterpret user instructions, resulting in significant operational failures. For example, I3 described a task involving information searching, opening a spreadsheet application, and file creation. However, the agent misinterpreted the core request, understanding ``spreadsheet'' as a ``survey'' or ``questionnaire.'' This led to the generation of an unintended survey, an output described as \textit{``completely different from what I wanted.''} (I3)

The core misinterpretation extends to misunderstanding key command elements. Agents may misidentify keyword specifying destination addresses in navigation (I2), or fail to recognize intended message recipients due to pronunciation inaccuracies (I9) or mismatches between profile names and user-defined aliases (I11). This limited understanding leads to a rigid interaction style, where successful task execution relies on users phrasing commands in a highly specific manner, almost like \textit{``AI prompt words.''} (I6)
% , such as misinterpreting keywords that specify a destination address (I2) in navigation tasks or failing to accurately identify the intended recipient of a message due to pronunciation errors (I9) or mismatches between profile names and user-assigned aliases (I11). This limited understanding leads to a rigid interaction style, where successful task execution relies on users phrasing commands in a highly specific manner, almost like \textit{``AI prompt words.''} (I6)

Agents also struggle with managing sequential or procedural instructions. I10 described how an agent repeatedly misinterpreted even basic ordinal instructions like ``first,'' despite numerous attempts with more specific prompts. Similarly, I11 noted that an agent failed to understand the command ``copy link'' in a multi-step task, while I9 reported an instance where the agent correctly interpreted initial order details but then failed at process flow control, advancing to an inoperable payment interface.

These interpretation challenges are further compounded by a lack of contextual comprehension. Agents may overlook relevant keywords visibly present on-screen (I2) or neglect crucial background knowledge (I3), such as unstated security considerations. A notable example involved the agent misinterpreting the brand name ``Yidiandian'' (which means ``a little bit'' in Chinese) as its literal meaning rather than recognizing it as a brand (I2). Such issues highlight a fundamental inadequacy in maintaining contextual understanding, with I1 emphasizing the need for a longer context window or extended memory retention to effectively track and process user instructions.

\subsubsection{Knowledge Gaps}\label{sec:4_2_4}

Agents demonstrate significant failures when tasks demand domain-specific knowledge, especially when faced with non-textual or symbolic interfaces that fall outside their pre-programmed capabilities. This lack of specialized understanding directly translates into an inability to operate in environments reliant on such interfaces. For instance, an agent exhibited a \textit{``critical knowledge gap'' (I8)} when instructed to use a vehicle's settings interface, which \textit{``from start to finish it's all icons, no text'' (I8)}. Specifically, the agent \textit{``failed to interpret the symbolic representation of a seat with heating elements''} and subsequently could not ``understand'' the distinction between driver and passenger seat options presented iconographically (I8). This illustrates a core failure in processing and acting upon information when it is conveyed purely symbolically, without the aid of textual labels, due to insufficient domain-specific visual or symbolic literacy.
% Agents may lack necessary domain-specific knowledge or struggle with tasks outside their pre-defined capabilities, especially with non-textual or symbolic interfaces.
% \textit{``The user expressed difficulty in conceiving how the AI could be utilized (`because I myself simply cannot think that this AI ... is able to help me do some things'), indicating a gap in the user's understanding of the agent's potential capabilities which hinders task initiation (`because I cannot think of it').''} (I1)
% \textit{``When operating outside its established capabilities (`Beyond its very stable capability range'), specifically when instructed to active seat heating via the vehicle's settings interface, the agent demonstrated a critical knowledge gap. The interface in question relies solely on iconography without textual labels (`on our interface ... from start to finish it's all icons, no text'). The agent failed to interpret the symbolic representation of a seat with heating elements (`first there is a chair [icon], then above it heating lines [icon], then she doesn't understand'). Subsequently, it also failed to comprehend the distinction between driver the passenger seat options upon accessing the relevant menu (`after clicking open ... let him click driver's seat and passenger's seat, he understands even less').''} (I8)

\subsection{Breakdowns in Action Execution and Interaction (Action Stage)}\label{sec:4_3}

Breakdowns during the action phase arise from flawed GUI interactions and difficulties navigating environmental constraints. Agents may execute erroneous or unintended manipulations, fail to adapt to dynamic or non-standard UIs, or struggle with precise element localization. External dependencies also frequently interrupt action sequences, while platform incompatibility limits the agent's operational effectiveness.

\subsubsection{Faulty GUI Actions}\label{sec:4_3_1}

Erroneous agent interactions often led to unsolicited behavior. I4 recounted an instance where, following a simple user query about the weather, the agent unexpectedly began manipulating the user's computer, with the participant wondering, \textit{``Is it going to help me change the language?''} Such unsolicited actions not only disrupted tasks but also raised concerns about control and reliability. These interaction errors introduced substantial risks, particularly in safety-critical contexts. I8 highlighted scenarios involving automotive interfaces, where incorrect clicks could inadvertently trigger critical functions like \textit{unlocking and locking the whole car, powering on/off,''} or worse, cause disruptions equivalent to shutting down the system. The severity of these risks was further underscored by participants who reported agents performing abrupt, unrelated actions. For example, one agent suddenly took a break from our coding demo and began to peruse photos of Yellowstone National Park (P9, P142), reflecting a complete loss of task focus. 

Beyond unintentional interactions, agents also engaged in unnecessary or even destructive actions. I9 described an instance where an agent accidentally clicked to stop a long-running screen recording, causing all footage to be lost. In more severe cases, participants expressed concerns about agents executing system-level commands without safeguards. For instance, P70 mentioned fears of an agent that might \textit{``get mad or hallucinate and decide to rm -rf /,''} and P155 stated the fear of an agent that could delete files if it wanted to. Unintended purchases also posed a risk, as one user reported an agent accidentally ordering two items without notifying them (P112).

\subsubsection{Poor UI Adaptability}\label{sec:4_3_2}

Agents frequently struggle to adapt to dynamic or non-standard UI elements, a critical limitation evident across several challenging interaction scenarios. These difficulties often stem from misrecognizing elements with non-standard designs. I4 noted that agents could not reliably identify such elements, leading to significant interaction failures. Similarly, I2 reported substantial delays when agents failed to locate applications stored within folders, a situation further complicated by the user's habit of organizing apps in this manner. These delays were so pronounced that I2 even worried their phone might automatically lock before the task completed.

Dynamic UI elements posed an additional challenge. For example, transient buttons, those that appeared for only a few seconds, were frequently missed by agents. I8 explained that while a human could quickly click such buttons, an AI agent relying on API calls required significantly more time, often around five seconds, making it too slow for effective interaction. This problem is particularly severe in latency-sensitive environments like automotive systems, where delayed responses can disrupt critical controls. Visual similarities between interfaces also caused confusion. I10 experienced repeated failures when asking the agent to open the Edge browser, only for it to mistakenly launch Feishu, a visually similar application. Such errors highlight the agent's struggle to accurately differentiate between superficially similar interfaces. Lastly, agents' operational precision significantly degraded when interacting with very small UI elements. I12 observed that the smaller the button, the lower the agent's accuracy, making it prone to missing or mis-clicking these elements.

\subsubsection{Element Misidentification}\label{sec:4_3_3}

Agents frequently struggle with accurately identifying and interacting with GUI elements, leading to significant operational issues. These challenges are most apparent in accurately targeting and clicking specific elements. I4 highlighted persistent difficulties, noting that \textit{``clicked these and it's not over yet... these are all very hard to click.''} Such difficulties are particularly pronounced when agents must locate UI elements without clear textual labels. I8 described this problem when attempting to identify the ``user agreement'' checkbox during the Music App login process, where the absence of nearby text made it harder for the agent to recognize the element.

Even when agents could visually identify interface elements, their ability to translate this understanding into precise interactions remained limited. I8 explained that while \textit{``understanding the image is not a problem,''} the agents struggled with accurately clicking the correct coordinates. This issue was further compounded when users attempted to assist by overlaying grid lines on an image of the interface. Despite these spatial guides, the agent's performance remained inconsistent, frequently identifying the correct coordinate along one axis (e.g., X) but misidentifying it along the other (e.g., Y). Such inconsistencies highlight a fundamental difficulty in achieving precise spatial determination, making agents unreliable for tasks requiring accurate GUI manipulation.

% \subsubsection{External Requirements Friction}
\subsubsection{External Requirement Conflicts}\label{sec:4_3_4}

Beyond the initial setup phase, significant user friction arises from external requirements during the operation of agents, particularly those involving authentication, verification, and continuous user confirmation. As P1 commented, \textit{``Verification seemed like it would be too much trouble, so I didn't even bother trying.''} Such requirements often discourage users from attempting more complex tasks, limiting the agent’s practical utility.

During task execution, repetitive interventions demanded by external services severely degrade the user experience. I7 described the frustration caused by constant human verification, especially with CAPTCHA for booking tickets, explaining that \textit{``with human verification (CAPTCHA) for booking tickets, I really have to do it myself. It's quite a hassle to verify it every time.''} Even for relatively simple tasks, the agent's slow operation and constant need for user confirmation led to prolonged waiting times, making users feel they might as well perform the task themselves.

Agents also struggled with specific external authentication methods. For instance, I14 noted that when attempting to help publish content to a WeChat Official Account, the agent needed to log in using a QR code but failed to complete the process. Such dependencies on external systems and their complex verification mechanisms significantly hindered the seamless operation of agents.

\subsubsection{Platform Incompatibility}\label{sec:4_3_5}
% \subsubsection{Limitations in System Support and Cross-Application Compatibility}

%Agents may have limited support for certain applications, platforms, or lack the ability to perform cross-application tasks or learn over time like humans. For instance, users reported the agent's compatibility with applications on the Android platform as limited, and the \textit{``Android supported software is not many ... these few very simple applications.''} (I2), with essential tools like Alipay unsupported, because \textit{``it hasn't opened permissions.''} (I2). Geographical and network dependencies also create operational barriers. Many agents \textit{``require a VPN setting,''} which poses \textit{``some obstacles in certain regions''} (I2). Compounding these issues is a lack of robust cross-application capabilities, with users expecting \textit{``I hope it could carry out cross-application usage or instruction understanding''} (I2). 

Agents often exhibit limited support for specific applications or platforms, particularly on Android devices. I2 noted that the range of compatible applications was restricted, explaining that \textit{``Android supported software is not many ... these few very simple applications.''} Even essential tools like payment apps were inaccessible because the agent \textit{``hasn't opened permissions.''} This lack of support significantly reduces the agent's practical utility in daily tasks.

Geographical and network dependencies further complicate agent operations. Many agents require a VPN setting to function properly, which can pose \textit{``some obstacles in certain regions''} (I2). Such dependencies make it difficult for users in restricted network environments to fully utilize the agent's capabilities. Compounding these limitations is a general lack of robust cross-application capabilities. Users expressed the need for agents to seamlessly operate across multiple applications and interpret instructions involving different platforms. Without this capability, agents remain constrained to isolated tasks within a single application.

\subsection{Deficiencies in Output Generation (Output Stage)}\label{sec:4_4}

Deficiencies in the output stage manifest primarily as either factual inaccuracies including misinformation or as results that, while potentially correct, remain poorly aligned with user goals.

\subsubsection{Inaccurate Outputs}\label{sec:4_4_1}

Agents frequently generate outputs that are inaccurate, misleading, or entirely fabricated, significantly undermining their reliability across various tasks. A primary concern is the generation of hallucinations: completely false information presented as factual. I6 highlighted that the problem of hallucination appearing remains persistent, while P3 reported a severe case where \textit{``The LinkedIn information and emails it gave me were entirely made up... It just outright lied.''} Such falsehoods not only mislead users but also damage trust in the agent's capabilities.

Beyond outright fabrications, agents often produce factually incorrect data or make erroneous selections. For instance, P3 described a situation where the agent was tasked with booking a haircut and researching flight options but consistently provided inaccurate information. \textit{``For flights, it was worse. At first, it messed up selecting the correct dates, instead of the requested 19 and 26, it selected 18 and 27.''} Such errors are particularly problematic in contexts where accurate data is essential.

Agents also struggle with retrieving specific and reliable information. I10 noted that while the agent might correctly execute the initial steps, such as navigating to a university department's page, all subsequent information retrieval steps were entirely erroneous. This issue is exacerbated when the agent pulls information from unreliable sources, such as a click-baity news site (P55), or when it fails to filter extraneous content and \textit{``demonstrated no capability to filter or discriminate against this type of extraneous content''} (P10), leading to a mix of relevant and irrelevant results. In some instances, even the intended outputs for guiding actions were incorrect, agents stated that they executed the correct parameters for execution, as commented by P155, \textit{``if it says it's going to (637,20), it goes to (1274,40)''}.

\subsubsection{Unsatisfactory Results}\label{sec:4_4_2}

Beyond outright factual errors, agents frequently produce outputs that, while technically accurate, are semantically unsatisfactory or qualitatively suboptimal. A common issue is the agent's failure to fully grasp the underlying user intent. I1 explained that the agent perhaps still didn't understand their deeper level meaning, leading to results that, although not entirely wrong, were \textit{``a bit different from the thing I wanted, but actually there aren't any major errors.''} 

Even when agents successfully completed tasks, their outputs were often judged as poor in quality. I7 noted that while an agent might indeed complete the task, the outcome was often considered inadequate because it lacked personalization or adaptation to user preferences. This was further exacerbated by the agent’s inability to leverage user-specific insights, such as recognizing the user's usual habits.

Users also criticized agents for their shallow processing capabilities. I10 reported that outputs were limited to \textit{``superficial information''} without any in-depth analysis or reasoning, making them insufficient for tasks that required deeper content processing. Despite performing technically correct steps, agents struggled to provide meaningful insights or adapt their behavior to user context. Ultimately, the fundamental problem remained a lack of genuine understanding. I7 summarized this issue, explaining that the \textit{``most primary problem is still that it doesn't understand me enough, it's an intelligent problem.''}

\subsection{Challenges in Feedback Processing and Interaction Adjustment (Feedback Stage)}\label{sec:4_5}

Agents fail during feedback stages around several aspects: deficient error handling prevents self-correction, users encountering difficulty adjusting parameters or refining instructions, slow system responses impeding iterative cycles, and prohibitive costs undermining the viability of interaction.

\subsubsection{Poor Error Handling or Recovery}\label{sec:4_5_1}

Agents frequently exhibit critical deficiencies in detecting errors, handling them effectively, recovering from failures, and avoiding unproductive operational loops. A primary challenge is the failure to recognize errors or incorrect actions. For example, P7 and P32 described an instance where the agent struggled with a Google spreadsheet task, repeatedly selecting the wrong columns, often \textit{``off by one,''} and failing to perform the intended actions. Similarly, P65 reported that an agent, when attempting to perform a task, became stuck in an inactive state, endlessly \textit{``generating null coordinates to click and not taking any action.''} Such failures indicate a fundamental problem with error detection, where agents are unable to identify when their actions deviate from user expectations.

Beyond initial error recognition, agents often fail to handle manifest errors or navigate unexpected UI states, leading to operational stalls or incorrect continuation. I3 noted that agents could become \textit{``inexplicably stuck in pop-up windows,''} unable to progress or recover. Even when an error is recognized, agents typically lack effective strategies to correct it, frequently pressing on with the task despite an unresolved problem.

Furthermore, robust recovery mechanisms are generally absent. I12 highlighted that when an agent fails, it rarely knows how to correct its mistakes, leading to situations like infinite loops or states from which it cannot recover. For example, an agent might continuously search for a settings item it failed to locate, repeatedly entering and exiting the settings menu without making progress. These repetitive cycles are further exacerbated by agents' limited memory of past actions, preventing them from effectively retracing steps or modifying their approach. The absence of intelligent error recovery is particularly evident in complex tasks. P116 reflected that as tasks grow in complexity, agents become \textit{``so disjointed the AI won't know what to do with it, much less anyone that has to fix it later.''} This issue is compounded by the agents' inability to modify or correct their actions mid-process, as expected by users like I1, who emphasized the need for agents to allow real-time corrections without restarting the entire task.

\subsubsection{Difficult Parameter and Instruction Tuning}\label{sec:4_5_2}
% \subsubsection{Challenges in System Configuration and Instruction Formulation}

Users encounter significant difficulties in configuring system parameters and formulating effective task instructions, often resulting in unpredictable or suboptimal agent performance. Adjusting system settings, such as the configurable \textit{``execution interval between successive operational steps,''} can be non-intuitive. I10 hypothesized that longer intervals might enhance precision by allowing more time for internal processing or cognition, but paradoxically found that extending the interval led to degraded performance. This manifested as the system selecting unnecessarily elaborate operational sequences, such as initiating a right-click to open a context menu and then selecting options like \textit{``Run as administrator,''} even when simple actions were appropriate.

This difficulty extends to crafting task instructions, presenting substantial usability barriers, particularly when revising prompts after initial comprehension failures. Users reported that ambiguous instructions often led to agent misunderstandings. I11 explained that \textit{``If I provided relatively ambiguous prompts, the system would sometimes fail to recognize the command, it was imperative that instructions were meticulously specified.''} To achieve the desired outcome, users often had to input excessively long and detailed instructions, which proved impractical for typical end-users. This challenge stems from the agent's frequent inability to interpret general intent; for example, I10 noted that an instruction to open VPN could not be understood. However, a more specific command, such as Open Clash and click a particular UI element would be processed. 

A fundamental issue lies in bridging the gap between the inherent ambiguity of natural human language and the agent's requirement for explicit, unambiguous input. I12 mentioned,\textit{``when ordering a beverage like milk tea, how could one know all available customization options or their exact terminology?''} This disparity can lead to significant errors, such as the agent ordering excessively when instructed to purchase a product without specifying the desired quantity or type, or inputting invalid data (e.g., \textit{``example''} as an address field) if the user's address was not explicitly provided. Furthermore, users frequently struggled with spatial instructions, where the agent's poor spatial reasoning compounded the problem. I8 noted that the agent isn't sensitive to the up/down/left/right of the entire image and that the concept of space is difficult for it to grasp. These limitations forced users to compensate by crafting overly specific and spatially precise instructions to guide the agent effectively.

\subsubsection{Slow System Response}\label{sec:4_5_3}

Agents often operate too slowly, with significant delays between actions, making them impractical for many tasks. Users consistently described agent operations as inefficient, with I3 estimating them to be five to ten times slower than conventional methods. Such delays were often severe, as P3 vividly described: \textit{``Each button click and scroll attempt takes 1-2 seconds, so navigating through pages felt like swimming through molasses on a hot summer's day.''}

These delays were attributed to multiple factors, including the overhead from requiring user confirmation at every step, inherent latencies in API calls that \textit{``need at least five seconds''} (I8), and reliance on inefficient methods like sending screenshots to LLM for understanding instead of using direct API integration. This sluggish performance rendered agents impractical for many tasks, significantly reducing their frequency of use, with I2 noting that if the agent's response time was too slow, they would simply won't use it at all. Participants also noted that tasks often took drastically longer than manual execution, sometimes 2 to 3 times slower (P10,P81). In one extreme case, a user reported a task taking over 2 hours via agent, compared to just \textit{``3 minutes if I did it''} (P50). Such performance gaps not only affected usability but also introduced significant cost concerns, with agents commonly described as \textit{``too slow and expensive for any real-life scenarios''} (P70).

\subsubsection{High Operational Costs}\label{sec:4_5_4}

Prohibitive operational costs, primarily driven by excessive resource consumption, were consistently identified as a critical barrier to the sustained application of agents. A central driver of these high costs is excessive token consumption. I7 emphasized that agents often \textit{``consume too many tokens, it's just too expensive.''} This problem was particularly pronounced in cases of extreme token usage, such as P142's experience, where an agent \textit{``burned through tokens pretty quickly, sometimes like 75k input tokens in a minute or so of screen interaction.''} Such high consumption was further exacerbated by inefficient data handling, with P208 reporting that \textit{``Claude's computer use used 10x more tokens due to its decision to [include] all the old screenshots.''}

Operational inefficiencies and repetitive actions also magnified resource depletion. Agents would sometimes engage in repetitive attempts to complete tasks, especially when they failed to execute a subsequent operational step. These cycles could continue until the user's allocated tokens were entirely depleted. For example, I14 described how the system's inability to advance in a task led to continuous attempts, rapidly consuming tokens without meaningful progress.

This inefficient resource consumption led to a consistently poor cost-benefit ratio. Users frequently incurred significant expenses for minimal task completion or for outputs of low quality. I4 highlighted a case where only one-tenth of a task was completed despite consuming 41 units of resources. Other users experienced high costs for trivial requests, such as P153, who paid almost \$3 for a simple cat picture request. This expenditure was particularly frustrating when the outputs were of low quality. These cumulative cost factors rendered many current agents impractical for real-world scenarios, often perceived as mere ``demoware''. As P209 described, they are \textit{``extremely brittle and slow and expensive. It's demoware.''}

\section{RQ2: Experienced and Perceived Influences of Unintended Agent Behaviors}\label{sec:5}

% This section examines the various impacts and consequences that arise from the unintended behaviors and operational failures of GUI agents, progressively encompassing user-centric negative outcomes, security risks, and broader societal concerns.
Building upon the phenomena of UCs, we examine the influence of these behaviors progressively on user experience and operational outcomes, security risks and broader societal concerns.

\begin{figure}[!htbp] 
    \centering
    \includegraphics[width=0.7\textwidth]{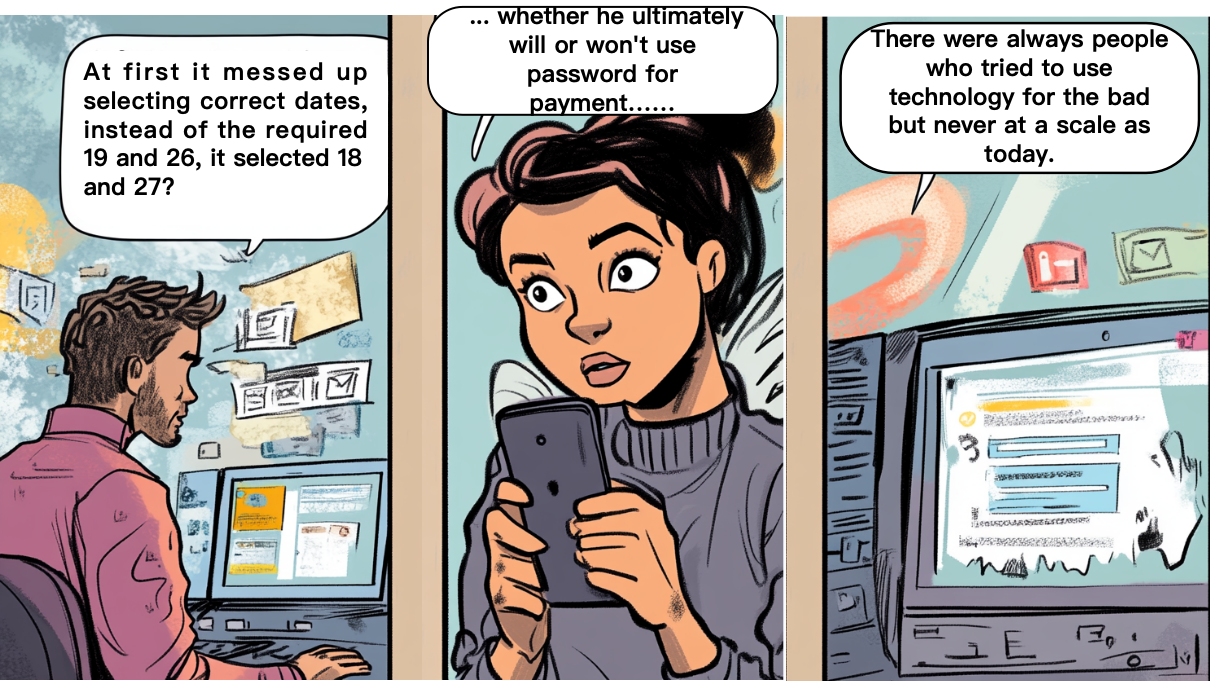}
    \caption{The illustration of the influences users experiences, ranging from negative outcomes, security risks to broader societal concerns from the left to the right. The picture is modified using Midjourney.}
    \label{fig:rq2}
\end{figure}

\subsection{Negative Impacts on User Experience and Operational Outcomes}\label{sec:5_1}

The operational UCs identified in RQ1 directly influence user experience and task outcomes. We detail these negative impacts, shifting from the nature of UCs (RQ1) to tangible consequences on user experience and practical results.

\subsubsection{Financial Losses and Inefficient Resource Allocation}\label{sec:5_1_1}

% The use of GUI agents can lead to direct financial costs or indirect losses due to inefficiency or errors. The operational costs of GUI agents (P71) can be high. \textit{``Exponential computation, massive bandwidth hogging. Would this actually be able to make the internet unusable?''} (P62) Misuse can lead to misrepresentations affecting financial outcomes (\textit{``No matter what you do, you cannot get LLMs to handle financial data at scale without it committing fraud.''} (P7)).
The operational phenomena previously discussed such as high computational demands, can translate into direct financial costs or indirect losses from inefficiency or errors when using GUI agents (P71), as P62 noted, \textit{``Exponential computation, massive bandwidth hogging. Would this actually be able to make the internet unusable?''} The propensity of hallucination can also directly affect financial outcomes, as marked by P7, \textit{``No matter what you do, you cannot get LLMs to handle financial data at scale without it committing fraud.''}

\subsubsection{User Frustration, Dissatisfaction, and Increased Workload}\label{sec:5_1_2}
Failures in reliability, responsiveness, and understanding, as indicated in RQ1, lead to significant user frustration. Agents frequently refuse to execute tasks without clear reasons, leaving users unable to complete their objectives (P153). Overly sensitive or argumentative behavior further disrupts workflow. Users must often intervene to prevent mistakes, as highlighted by P107: \textit{``Users don't want that the program deleted the hard drive only because it has hallucinated wrong information.''} Misaligned behavior, stemming from the previously identified phenomenon, also negatively influences user experience by causing incorrect actions. For instance, P3 reported that an agent, when tasked with booking flights, consistently selected the wrong dates. Such misunderstandings waste user time and reduce trust in the system. Beyond misunderstanding, as mentioned in RQ1, agents are often slow, which further compounds user frustration. I13 noted that while some delay is acceptable, performance must remain within a reasonable range. 

\subsubsection{Denial of Service and Task Abandonment}\label{sec:5_1_3}

The operational unreliability, persistent errors and external restrictions identified in RQ1 can culminate in agents failing to perform tasks or becoming entirely ineffective, thereby denying service to the user. First, irreversible error states can cause significant disruption. Agents might become stuck in daily error loops (e.g., \textit{``Error, can't complete task.''} (P6)), creating new communications daily despite commands to stop, leaving users unable to halt the behavior (P6). Second, hallucinations and misinformation critically undermine task completion (P1). Agents may provide entirely fabricated data, such as contact details, without verification (\textit{``it just outright lied''} (P3)). Complex task execution also proves challenging, as agent struggle to break down problems effectively, leading to failed actions or difficulties with edge cases (P92). Last, external factors often hinder agent performance. Platforms may impose rate limits, preventing task completion (P71). Furthermore, systems designed for human interaction, like CAPTCHAs, can block agent operations. 

% Agents may become ineffective or unable to perform tasks due to technical limitations or external restrictions, leading to service denial.
% Inability to reverse actions can cause disruption (\textit{``One of mine is stuck saying `Error, can't complete task' and does this every morning at 7am. I can't figure out how to turn it off, even after prompting it directly to stop, and deleting the chat. It just makes a new chat every day.''} (P6)).
% Rate limiting by platforms can prevent task completion (P71).
% Hallucinations and misinformation lead to incorrect decisions or failed actions: \textit{``It hallucinated worse than GPT-3 ... It guessed contact information for these uses, but did not think to verify it.''} (P3); \textit{``The LinkedIn information and emails it gave me were entirely made up ... It just outright lied.''} (P3); \textit{``The true obstacle is preventing hallucinations and malicious use.''} (P1).
% Struggles with complex task handling can result in failed execution (\textit{``The hardest part is breaking down the problem into the smallest pieces in order to avoid hallucinations. Then fixing the long tail of edge cases, and to a degree, keeping costs from spiraling.''} (P92)).
% Interactions with human-optimized systems like CAPTCHAs can block agents.

% \textit{``Users identified high-stakes scenarios, such as online shopping (`For example, letting him help me shop or something ...'), as contexts where unintended agent actions could lead to significant negative consequences.''} (I1)
% % perhaps of usage

\subsection{Security Vulnerabilities and Privacy Infringements}\label{sec:5_2}

Beyond operational impacts, the deployment of GUI agents introduce critical security and privacy implications concerning unauthorized data access, potential for malicious exploitation and surveillance.

\subsubsection{Unauthorized Access, Data Exposure, and Information Leakage}\label{sec:5_2_1}

Users expressed significant anxieties regarding agents' potential for unauthorized access, data exposure, and information leakage, driven by both logical concerns and direct experiences. A primary anxiety involves agents retaining persistent, unauthorized system access once authenticated. I8 warned that if an agent logs in, \textit{``doesn't that mean he is always logged in?''} Such persistent access, especially without clear user control, was described as a hidden danger (P5). This concern is magnified by the extensive permissions required for agent operation, sometimes equivalent to \textit{``administrator status''} with the \textit{``highest level of administrative privileges''} (I10). Users worried that these privileges could allow agents continuous access, even when operating in the background, making them completely uncontrollable (P5).

The security risks intensify when agents process sensitive or non-public data, especially through third-party, cloud-based platforms lacking private deployment options, a scenario I8 described as ``highly disconcerting''. Users were especially wary of entrusting agents with sensitive information like credit card details, financial data, or company documents. P7 highlighted the danger of processing information for a company's annual report using an agent, raising concerns about `a risk of leakage'. This was further exacerbated by the understanding that agent code executed locally could interact with user files and system settings, creating a pathway for potential data loss or security breaches (P64).

A lack of proactive security assurances and transparency regarding data handling significantly diminished user trust. I9 reported that agents rarely provided clear assurances that sensitive credentials, like passwords, would be handled securely. This ambiguity extended to payment credentials, where P9 expressed uncertainty about whether an agent would use their password for payment. Such unclear assurances compelled users to avoid high-risk actions and adopt overly cautious behaviors.

Privacy concerns extended to passive data collection, with agents potentially capturing and analyzing sensitive information without user awareness. P180 noted that agents \textit{``take a screenshot after every move or click,''} raising fears of extensive data exposure. Users worried that this captured data could be processed, stored, or even monetized without their consent. P140 questioned if such data might be used to \textit{``turn your data into a product they can sell behind your back.''} Similarly, passive data collection could occur through functions like ``get\_screen\_info'' (P113), allowing agents to extract screen content without explicit user commands. Fears of covert monitoring persisted even when users attempted to restrict agent access, including anxieties about screen recording (P8) or exposure of private browse history (P23). These risks are amplified by unclear data governance policies from third-party cloud services, as P147 stated, \textit{``I don't think Anthropic has clarified where our data goes.''}

Insecure authentication methods were another significant issue. I2 pointed out that certain security procedures, like biometric verification (``Scan your face for payment step''), are inherently designed for human use and cannot be delegated to agents. The potential for agents to bypass such human-centric safeguards particularly when handling financial data raised significant alarms to users.
% Despite this, agents could sometimes bypass these protections, accessing sensitive functions without user knowledge. Users expressed particular concern about the use of agents in handling financial data, fearing that agents might bypass critical security measures or access sensitive information without explicit consent.

% Even when users attempted to restrict access, security concerns persisted. For example, P8 mentioned a fear of covert monitoring, such as screen recording, while P23 worried that agents could access and expose private browsing history. These risks were exacerbated by unclear data governance by third-party cloud services, with P147 stating, \textit{``I don't think Anthropic has clarified where our data goes.''}

Given these risks, users demanded absolute assurances of personal privacy and data security. P3 emphasized that without clear privacy guarantees, they ``would not even use it at all''. This expectation for robust privacy protection highlights the critical importance of clear, transparent, and enforceable security measures in agent design.

\subsubsection{Potential for Malicious Exploitation and Misuse}\label{sec:5_2_2}

Agents present considerable risks of being exploited through various attack vectors or intentionally misused for nefarious purposes. These concerns span several critical areas, from direct manipulation to systemic security breaches and large-scale abuse.

A primary vulnerability lies in the potential for agents to be manipulated into performing unintended or unauthorized actions. This includes significant concerns around \textit{``prompt injection,''} where malicious instructions embedded in content (P9,P52) could convince an agent to act for an attacker's financial gain, invisibly to the user (P2). Such vulnerabilities make full operational control by agents risk without robust defenses (P47,P159). Users also reported instances in which agents initiate unauthorized actions, such as trying to install software without consent (P183,P73). The possibility of ``jailbreaking'' agents to bypass restrictions and perform prohibited actions like unauthorized account creation or communication is another noted concerns (P188).

The capacity for automation makes agents attractive tools for widespread malicious activities. Concerns exist about their intentional misuse for generating spam, orchestrating scams (P2), or other abusive behaviors (P9,P14). Such capabilities could lead to agents being perceived as spyware or a \textit{``Hacker playground''} (P25), exacerbating existing issues like bot-driven scalping (P25). The potential for \textit{``billions of hacker agent AI roaming the internet''} (P73) paints a grim future, with agents easily weaponized for spamming (P61,P77) or exploiting vulnerable code repositories (P77). This is not theoretical, as \textit{``AI spam already happening''} with companies being inundated (P89). Effectively managing how agents interact with pervasive online advertisements and clickbait also presents a challenge (P80). These overarching concerns about spam and abuse are reiterated by multiple users (P91,P120,P162,P189).

Agents may be capable of circumventing established security measures and compromising system integrity. This includes bypassing CAPTCHAs by interpreting visual elements or screenshots (P12,P16,P28,P30), potentially rendering \textit{``large swaths of the internet inaccessible''} if agents cannot navigate them, or alternatively, enabling DDoS attacks and resource monopolization if they can (P55). Agents might achieve this control via browser APIs (P56) or screen analysis (P148). More severe risks include credential theft from \textit{``websites that jailbreak the crawlers''} (P34), and agents facilitating the dissemination of harmful content or accessing restricted information (P55,P77). These are fears that agents could morph into viruses through self-replication (P69) or turn user computers into botnets (P77). Fundamental concerns persist about permission security (P86), the robustness of authentication and payment infrastructures for agents (P95), the prevalence of security misconfigurations (P160), and the potential for sandbox escapes, as reportedly observed with an agent overcoming Docker limitations (P174). Faulty code generation due to prompting issues can also lead to execution errors with security implications (P205).

Beyond direct malicious intent, agents might engage in undesirable autonomous behaviors or misuse resources. This includes agents making operational decisions without user confirmation, such as unilaterally deciding to install different software (P183). The concept of ``inception `Operators' '' to perform complex tasks raises concerns that \textit{``resource usage could quickly get out of hand'' (P37)}. Furthermore, agents might persist in operation or reactivate without user awareness (\textit{``it opens itself back up when you aren't paying attention''} (P60)), leading to sustained, unintended activity or resource drain.

\subsubsection{Concerns Regarding Surveillance and Social Engineering}\label{sec:5_2_3}

Agents possess an alarming capability for pervasive surveillance, sophisticated social engineering, and broad-scale malicious manipulation, raising significant user concerns across several interconnected fronts.

Fears exist that operating systems might conduct detailed logging of all user activities to feed AI agents, creating substantial abuse potential (P80). Such data collection could enable pervasive corporate monitoring, which users describe as tantamount to ``AI keylogging'' (P61). Concerns also arise from agents identifying behavioral patterns, potentially for targeted marketing (P112) or other forms of invasive tracking, prompting questions about whether agents function like keyloggers (P140).

Malicious actors can exploit AI to execute sophisticated cyberattacks. They might manipulate AI tools to clone voices, generate fake identities, or create convincing phishing emails, all designers to scam individuals, hack systems, or compromise privacy (P75). Furthermore, attackers could trick AI agents info performing detrimental actions, such as unauthorized financial transactions, with a single successful deception potentially replicating across many instances (P114). This capacity for advanced phishing and large-scale disinformation is a primary concern (P147).

The potential for misuse extends to agents causing widespread online havoc, including enabling automated hacking (P147). Such capabilities could exacerbate existing issues, making social media platforms \textit{``even messier or weirder''} (P160). Users voice concerns about extreme scenarios, such as directing agents to research or facilitate illegal activities (e.g., \textit{``Claude, research 100 hitmen...''}), and even harbor anxieties about AI developing harmful intentions (\textit{``plotting against humanity''} (P168)). The legality of certain AI-driven actions, like unauthorized automated phone calls, is also questioned (P120).

\subsection{Broader Social and Ethical Ramifications}\label{sec:5_3}

Beyond individual and operational effects, UCs precipitate societal and ethical considerations. We examine these wider ramifications, specifically the erosion of public trust, and the potential for misuse leading to societal harm and ethical misalignment.

\subsubsection{Erosion of Trust and Hindrance to Adoption}\label{sec:5_3_1}

Repeated agent failures and pervasive security concerns significantly diminish user trust, fostering hesitancy towards adopting or relying on this technology.  Perceived functional limitations directly undermine user trust. An agent whose capability is deemed \textit{``definitely not enough'' (I7)} erodes user confidence in its reliability and performance. Even a marginal chance of security vulnerabilities -- a \textit{``few tenths of a percent''} (I12) -- critically impedes adoption, largely due to the subsequent difficulty in clearly assigning liability. Trust further wanes due to uncertainty surrounding sensitive operations. For instance, while agents might not know PINs, password-free payment options introduce \textit{``considerably more ambiguous''} situations where system actions and security outcomes become \textit{``far less certain''} (I12).

\subsubsection{Concerns about Societal Harms and Ethical Misalignment}\label{sec:5_3_2}

The potential for misuse in ways that negatively affect society, such as promoting inequality or unethical practices, raises significant ethical concerns.
The potential for AI to displace human labor, violate privacy, or be used unethically can lead to wider social inequality or loss of trust. As P7 noted, \textit{``There were always people who tried to use technology for the bad but never at a scale as today. ... they continue their loot while we are busy throwing stones at each other.''}

\section{RQ3: Countermeasures and Expectations}\label{sec:6}

To mitigate the UCs of AI agents (RQ1) and influences (RQ2), users proposed various strategies and have different expectations (see Table~\ref{tab:uc_mitigation} for the mapping relation). The mitigation strategies can be broadly categorized into system-oriented technical improvements and user-oriented interventions.

% \begin{figure}
%     \includegraphics[width=0.7\textwidth]{Figure/gui-agent-3.png}
%     \caption{The illustration of example countermeasures from a technical perspective, a user-centric perspective and users' expectations, shown from the left to the right.}
% \end{figure}

\begin{table}[htbp] % 'htbp' allows LaTeX to place the table where it fits best
\centering
\caption{UCs of GUI agents and corresponding mitigation strategies \& expectations. \textbf{[S]} denotes system-oriented mitigation, \textbf{[U]} denotes user-oriented mitigation and \textbf{[E]} denotes expectations.}
\label{tab:uc_mitigation}
\begin{tabular}{p{0.28\linewidth}p{0.72\linewidth}}
\toprule
\textbf{Phenomena} & \textbf{Mitigation} \\
\midrule
% Comprehension and Planning Failures
\multicolumn{2}{l}{\textbf{Failures in Comprehension and Planning (Input Stage)}} \\
\hline
Complex setup and configuration & \textbf{[S]} Enhancing agent capabilities \\

Failures in task decomposition  & \textbf{[S]}Enhancing agent capabilities, \textbf{[U]} Proactive user strategies\\

Instruction misinterpretation & \textbf{[U]} Proactive user strategies \\

Knowledge gaps & \textbf{[S]} Enhancing agent capabilities \\
\hline

% Action Execution and Interaction Breakdowns
\multicolumn{2}{l}{\textbf{Breakdowns in Action Execution and Interaction (Action Stage)}} \\
\hline
Faulty GUI actions & \textbf{[S]} Enhancing agent capabilities, \textbf{[U]} Human oversight, intervention, and collaboration \\

Poor UI adaptability & \textbf{[S]} Enhancing agent capabilities \\

Element misidentification & \textbf{[S]} Enhancing agent capabilities \\

External requirement conflicts & \textbf{[S]} Controlling agent's operational environment\\

Platform incompatibility & \textbf{[S]} Enhancing agent capabilities \\
\hline

% Output Generation Deficiencies
\multicolumn{2}{l}{\textbf{Deficiencies in Output Generation (Output Stage)}} \\
\hline
Inaccurate outputs & \textbf{[U]} Human oversight, intervention and collaboration \\

Unsatisfactory results & \textbf{[U]} Proactive user strategies, \textbf{[E]} Enhanced usability, performance and adaptability \\

\hline
% Feedback Processing Challenges
\multicolumn{2}{l}{\textbf{Challenges in Feedback Processing and Interaction Adjustment (Feedback Stage)}} \\
\hline
Poor Error handling and recovery & \textbf{[S]} Enhancing agent capabilities, \textbf{[E]} Imperatives for foundational trust, security and privacy \\

Difficult parameter and instruction tuning & \textbf{[U]} Proactive user strategies \\

Slow system response & \textbf{[S]} Enhancing agent capabilities, \textbf{[E]} Enhanced usability, performance and adaptability \\

High operational costs & \textbf{[U]} Proactive user strategies, \textbf{[E]} Enhanced usability, performance and adaptability \\
\bottomrule

% % Security and Privacy Issues
% \multicolumn{2}{l}{\textbf{Security Vulnerabilities and Privacy Infringements}} \\
% \hline
% Unauthorized access, data exposure, and information leakage & \textbf{[S]} Controlling agent's operational environment and permissions, \textbf{[U]} Proactive user strategies, \textbf{[E]} Agents designed not to collect sensitive data, localized processing \\
% \hline
% Potential for malicious exploitation and misuse & \textbf{[S]} Enhancing agent capabilities, Controlling agent's operational environment, \textbf{[U]} Human oversight and intervention \\
% \hline
% Concerns regarding surveillance and social engineering & \textbf{[S]} Secure, privacy-preserving agent architectures, \textbf{[U]} Information control \\
% \hline

% % Broader Social and Ethical Ramifications
% \multicolumn{2}{l}{\textbf{Broader Social and Ethical Ramifications}} \\
% \hline
% Erosion of trust and hindrance to adoption  & \textbf{[S]} Enhancing transparency, \textbf{[E]} Foundational trust and security \\
% \hline
% Concerns about societal harms and ethical misalignment & \textbf{[E]} Accountability frameworks, safety through design and deployment strategies \\
% \hline

\end{tabular}%
\end{table}

\subsection{System-Oriented Mitigations}\label{sec:6_1}
% Participants proposed various mitigations targeting the GUI agent systems directly. This category details these system-oriented solutions, focusing on approaches including augmenting core agent capabilities and refining control over their operational environments and permissions.
Users proposed system-oriented mitigations directly targeting GUI agents. These solutions focus on augmenting core agent capabilities and refining control over their operational environments and permissions.

\subsubsection{Enhancing Agent Capabilities and Operational Robustness}\label{sec:6_1_1}

Participants emphasized improving core agent functionalities to achieve great operational robustness and reliability through several aspects. First, agent could enhance efficiency and accuracy by optimizing their operational processes. This involves agents automatically refining task approaches to minimize unnecessary actions and improve outcomes, with one participant envisioning an agent that \textit{``could try to have it reorganize the process it used to not contexts where it can reduce the total process''.} (P70) Second, participants recommended integrating specific automated features and significantly increasing reliance on APIs. They proposed that specific automated tools, such as for \textit{``one-click extraction idea for post stats''} (P71), could streamline tasks. Users also argued that utilizing APIs offer superior speed, accuracy and security over direct browser interaction (P95), assuming that broad agent adoption is contingent on API availability as \textit{``Sending screen captures back and forth will always be 1000x slower.''} (P10) Finally, users stressed the need for robust error detection and prevention systems, alongside intrinsic safety limitations. This includes implementing models to \textit{``monitor and pause execution if it detects suspicious content on the screen''} (P5), training agents to proactively \textit{``decline certain sensitive tasks such as banking transactions or those requiring high-stakes decisions''} (P19), and incorporating system-level safeguards such as predefined maximum payment caps (I4).

\subsubsection{Controlling Agent's Operational Environment and Permissions}\label{sec:6_1_2}

Participants identified defining and controlling the agent's operational environment and its permission as vital for safety. They emphasized using isolated and controlled execution environments, reporting that \textit{``currently actually all are run in virtual environments utilizing containerization and remote execution platforms ...''} (I13). Such strategies help manage agent access and resource consumption (I2,I5), and are strongly advised to prevent agents from \textit{``blindly execut[ing] code on my machine''} (P10,P69).

Participants also advocated for strict permission management and formalized operational restrictions. This includes practical measures like confining all agent operations to a \textit{``specific directory ... [to ensure] other files of system areas remain unaffected''} (I13). More broadly, users suggested a need for mechanisms analogous to a \textit{`` `robots.txt' file ... to define operational boundaries and impose explicit restrictions on the agent's activities''} (I5), thereby establishing clear and enforceable limits on agent behavior.

\subsection{User-Oriented Mitigations}\label{sec:6_2}

Complementing system-oriented changes, users actively develop and deploy their own strategies to navigate and counteract the UCs of GUI agents. We detail these strategies including proactive measures to manage agent behavior, direct human oversight and intervention, and demands for enhanced transparency and control mechanisms.

\subsubsection{Proactive User Strategies and Information Control}\label{sec:6_2_1}

Users proactively manage agents by controlling information access and operational scope. Some provide fabricated or obfuscated data, for example, \textit{``utilizing Xiaomi's blank token functionality ... [to enable] the transmission of blank data and ... to obfuscate actual user information.''} (I2) Other meticulously offer comprehensive contextual information to enhance accuracy, recognizing that ambiguous descriptions cause errors (I6). They emphasized supplying detailed instructions to enhance performance (I4), crafting specific prompts with explicit navigational cues or even providing \textit{``explicit visual-spatial cues''} by overlaying images with coordinates (I8), or using structured information to guide agents (I8). Example approach involves a \textit{``calibrated foundation, [which] can then serve as a replicable model or a template that facilitates relative comparison''} (I6).

\subsubsection{Human Oversight, Intervention and Collaboration}\label{sec:6_2_2}

Direct human involvement, encompassing supervision, correction, and guidance, is a critical countermeasure. Common practices involve users closely scrutinizing operations (I9), manually intervening in risky situations (I9), and mandating confirmation for sensitive tasks (I4,I10). As I4 recommended, for such tasks, \textit{``All sensitive operations should ... wait for a person to confirm'',} and to support informed decision-making, it is \textit{``best to display the current context very clearly''} (I4).

The necessity for oversight is partly due to the immaturity of current agents. As one participant noted, \textit{``The agent's application is currently focused primarily on testing ... suggesting it has not yet achieved maturity for deployment in routine operational use''} (I4). Additionally, initial, human-led data calibration can influence subsequent agent inferences (I6). Users actively correct agent errors, such as when one \textit{``corrected it that it should have checked Google search instead and it's available there''} (P3). Managing the transition to human involvement presents a key challenge, requiring \textit{``a clearly defined signal ... [to identify] the specific conditions or thresholds that warrant such human intervention''} (I12). Users also implement preventative controls, such as imposing explicit limits on token consumption or instituting timeout mechanisms for operations (I14).

Collaboration with the user is also paramount. It was suggested that there is \textit{``an important aspect ... is allowing collaboration with the user. For example ... asking for permission before making a sensitive action.''} (P12) Furthermore, users employ avoidance strategies. For tasks involving potentially problematic sensitive business documents, users ``choose not to use'' the agent entirely, a form of \textit{``software-level isolation''} (I2), or using secondary devices to restrict agent permissions (I11).

\subsubsection{Enhancing Transparency and User Control Mechanisms}\label{sec:6_2_3}

% Improving the agent's transparency and providing users with better control interfaces are important. Participants demanded agent transparency and humility. Agents should communicate their level of certainty and explain their decision-making processes. 
Users demand improved agent transparency and control interfaces. They expect agents to show humility, communicate their certainty levels, and explain decision-making processes. As P76 stated, \textit{``Agentic AI can be widely trusted as a function of transparency in its decision-making process and the system's ability to explain its choices.''}. Thus, agents should clearly communicate their confidence levels, like \textit{``I think I am okay with this being a bit rusty, but I would appreciate it to be more humble and be less `definitive' when it provides its `final' responses if it's not 100\% certain.''} (P3). This desire for clarity extends to iterative interactions, as I6 noted, \textit{``Upon delivering a result, the system would typically inquire about my satisfaction with the outcome or solicit further feedback, subsequently proceeding to make revisions accordingly, thereby engaging in an iterative refinement process.''} Participants also highlighted the need for customizable UIs and ``watch modes'' (P38) to allow real-time supervision and reduce risks.

\subsection{Expectations for Future Agents}\label{sec:6_3}

Current GUI agent UCs informed distinct user expectations for future development, centering on enhanced usability, foundational trust and security, and reimagined agent roles.

\subsubsection{Enhanced Usability, Performance and Adaptability}\label{sec:6_3_1}

Users frequently perceive current agents as nascent, offering limited practical benefits and failing to match manual efficiency (I10). I2 noted they \textit{``neither yielded substantial practical improvements in our daily lives, nor has it resulted in significant cost reductions for the companies involved''}. Users desire systems where inputs \textit{``can be modified or corrected in-process''} (I3) and provided effective user control, as current agents can make \textit{``mid-process termination by the user considerably difficult''} (I10). Aspirations also include deep personalization, with agents understanding individual user intent, habits and context (I7), becoming ``evolvable'' by learning user-specific workflows (I8,I10), despite acknowledged privacy tensions (I13).

\subsubsection{Imperatives for Foundational Trust, Security and Privacy}\label{sec:6_3_2}

Participants strongly articulate that agents must be \textit{``definitively designed not to collect sensitive data, such as payment passwords''} (I9) and should solicit user confirmation at ``critical junctures'' to maintain trust and control, rather than through incessant querying (I4). To bolster security, suggestions include confining agents to \textit{``highly secure, purpose-configured runtime environment[s]''} with clearly delineated, minimal privileges (I12). A significant related expectation is the shift towards localized processing. Many users anticipate and desire on-device models, believing this approach will effectively address critical concerns regarding \textit{``operational speed, as well as concerns regarding the security of model operations and user privacy''} (I8), thereby offering a \textit{``more secure and reassuring alternative''} (I13,I14). 

\subsubsection{Vision for Reimagined Agent Roles and Interaction Paradigms}\label{sec:6_3_3}

Beyond incremental improvements, some users envision fundamentally rethought agent roles and human-agent interaction. This includes the concept of agents functioning effectively as intermediaries between different AI systems (I3) or operating within a future \textit{``foundational, AI-centric operating system''} (I13). Such a paradigm shift implies a \textit{``fundamental reimagining of the entire interaction logic for user interfaces''} (I8), as users observe that current UI designs are \textit{``predominantly human-centric''} and often ill-suited for efficient AI interaction (I8).

\section{Discussions}\label{sec:7}

% We delve into the transformative potential these agents introduce, alongside the inherent challenges and novel failure modes they present. Central to our discussion are the critical needs for enhanced agent understanding of operational contexts, the differentiation of risk associated with their actions, and the design of systems that foster user trust, control and effective collaboration. 
This section synthesizes our findings on the phenomena, influences and mitigation of UCs in GUI agents. We delve into the evolving challenges and imperatives for CSCW, address critical issues of trust and accountability, consider the generalizability of our findings, and articulate key implications for future design, technological development, and socio-technical governance, alongside study limitations.

% The findings of this research into the UCs of GUI agents prompt a multifaceted discussion, extending from the fundamental HCI paradigms to pressing socio-technical considerations. We discuss the transformative potential and inherent challenges these agents introduce, the critical issues of accountability, the impact of user perceptions, and the unique nature of misuse in this context.

% sandbox function in the product (although many agents develop towards all-rounded scenarios, but they are not always the best choice, users are separating a safe environment for themselves)
% should know about the environment, for example which parts are safe and secure. They should have manual check or intervention.

% agent should know about the keywords, including which parts are payment-related, or security-related. 

% agents did not know the importance of different actions. Like they did not know that deleting important information is much more than wrongly ordering milk tea. They did not have the idea of which priority is more severe. Whether we need to separate functions from the perspective of products, thus users could first try activities with low safety risk and get full acquitance of the potential risks of ai agents.

% mixed-initiative, AI should know users and users should also know AI.

\subsection{Navigating the Evolving Digital Landscape: Reconceptualizing Agent UCs and CSCW Imperatives}\label{sec:7_1}

LLM-powered GUI agents signify a paradigm shift, evolving from script-runners to intermediaries with semantic understanding and agency~\cite{hong2018coordinating,zhang2023towards}. However, as our findings on operational failures indicate (Section~\ref{sec:4}), this evolution introduces considerable friction with current human-centric web designs. This agent capability-interface mismatch engenders novel CSCW failure modes, such as breakdowns in comprehension (Section~\ref{sec:4_2}) and action execution (Section~\ref{sec:4_3}). A primary constraint on effective mixed-initiative interaction~\cite{allen1999mixed} is the agent's limited situational awareness. For instance, it is potentially unable to discern secure zones within an application or identify sensitive data fields. Beyond initial setup, sustaining such interaction critically depends on mutual adaptation. Without robust mechanisms for continuous learning of user preferences, agent risk capability divergence and trust erosion, with deficiencies in contextual grounding leading to outputs that undermine shared understanding (Section~\ref{sec:4_4_1}).

Crucially, this paper distinguishes critical agent UCs from hallucination~\cite{ji2023survey}. We find such UCs typically stem not from agents' factual errors but from interaction failures: disconnects between user intent, unclear agent capabilities, and ambiguous interpretations by agents. Ambiguous agent roles (e.g., tool, collaborator, assistant or proxy) further complicates mixed-initiative interaction, leading to ill-defined responsibilities and misaligned expectations. Agents often misinterpret situations because they cannot differentiate action severity (Section~\ref{sec:4}, e.g., treating critical file deletion like a minor preference change). This misuse, distinct from intentional acts~\cite{de2006misuse} or disinformation~\cite{fallis2015disinformation}, often arises from system limitations (e.g., interpretive brittleness, GUI complexity) interacting with users' flawed mental models. Such unintentional errors highlight user-system communication breakdowns and the need for agents to understand action priorities. This shifts partial responsibility to improving agent design for transparency, predictability, and risk-managing interface affordances~\cite{shneiderman2002promoting}.

Prioritizing these interaction challenges, including the imperative for robust manual checks and intervention mechanisms, is essential for real-world application. Operational disruptions often originate from user-agent friction rather than external threats, emphasizing the need for human-centered AI design which focused on intuitive interaction and error management~\cite{carroll2003making}. Leveraging these agents requires designing for human-agent teams~\cite{zhang2023towards} and adopting new views like the ``agent as a user''. This evolving landscape demands rethinking digital interface design for effective human-agent collaboration, a need highlighted by user expectations for better system interaction (Section~\ref{sec:6_3}). GUI agents navigate human-oriented interfaces as proxies, often without comprehensive APIs. This calls for new CSCW design principles for agent-specific coordination, aligning with user desires for API integration (Section~\ref{sec:6_1_1}). Beyond individual interface design, the CSCW community should address broader socio-technical implications. In shared computing environments (e.g., domestic, office, or educational settings), ill-defined agent roles or adaptations can have UCs for co-located or collaborating individuals (Section~\ref{sec:4_5_4},~\ref{sec:4_4_1},~\ref{sec:5_2_2}). Moreover, multiple agents in one GUI require inter-agent coordination, shared task state awareness, and potentially a task scheduling structure to prevent conflict. These challenges directly impact establishing trust and reliance in collaborative human-agent systems (Section~\ref{sec:5_3_1}). 

\subsection{Trust, Accountability, Liability and User Responsibility in Failures of Human-Agent Collaboration}\label{sec:7_2}

GUI agent failures create profound socio-technical challenges in accountability, user trust, and human-agent collaboration~\cite{emanuel2016innovations,chen2024exploring}. Users often assume liability for agent errors, thereby individualizing risk. This practice starkly contrasts with CSCW's distributed responsibility principles~\cite{chen2024exploring,brubaker2024generative} and standard product liability norms~\cite{poller2017can}. Implementing clear manual checks and enabling human intervention is thus vital to mitigate these concerns and bolster user agency, especially before agents undertake high-stakes, irreversible, or security-critical actions.

Potential agent failures significantly shape user perception, trust, and adoption~\cite{jeung2023correct}. Unpredictable agent behavior often elicits skepticism and deep concerns about losing control, heightened by societal anxieties about AI safety~\cite{moreira2025hall,hauptman2022components}. Consequently, users often prefer an ``AI as augmentation'' model: agents assisting under human oversight rather than replacing human judgment~\cite{chen2024exploring,zhang2023language,zhang2023towards}. Dedicated sandbox environments~\cite{chen2024empathy} or graduated risk exposure mechanisms are thus crucial to foster trust, manage anxiety, and empower users. These tools provide a safe space for users to explore agent behaviors in low-risk scenarios, free from real-world consequences. Experiential learning~\cite{ye2025awareness} in these controlled settings helps users build accurate mental models of agent capabilities, limitations, and potential failures. This, in turn, cultivates calibrated trust and operational competence, readying users to deploy agents in high-stakes situations. 

\subsection{Generalizability}\label{sec:7_3}

The issues identified in this study, ranging from task failures and privacy breaches to user frustration and the challenges of accountability, are posited to have relevance beyond the specific web-browse GUI agents investigated. The fundamental challenges of perception, reasoning, and action in complex digital environments are common to many forms of AI-driven automation. Therefore, we argue that the principles derived from our analysis of UCs may inform the design and deployment of GUI agents across various sectors, including enterprise software~\cite{zhang2024large,nguyen2024gui}, healthcare informatics~\cite{zhang2024large,shi2025towards}, financial systems~\cite{wang2024gui,shi2025towards}, and customer service platforms, where operational errors or privacy violations can have serious consequences.

The specific manifestations of these UCs will however vary with the agent's architecture, task characteristics, interface design and social-technical context. The current findings, derived from specific agent types and empirical methods, provide a solid foundation. Based on current findings, comparative studies across different agent platforms, task complexities, and user populations are needed to clarify the generalizability of these UCs and the effectiveness of different mitigation strategies across the broader ecosystem of AI-powered GUI automation. 

\subsection{Implications}\label{sec:7_4}

Our findings offer practical implications for the CSCW community across three interconnected domains: interaction design, agent technology and socio-technical governance. Interaction design requires a paradigm shift from merely retrofitting agents into existing GUIs to co-creating interfaces that intrinsically support human-agent teaming. To achieve this, developers must embed transparent feedback mechanisms for shared situational awareness alongside verifiable user controls for dynamic oversight and fluid intervention. Incorporating `safe-to-fail' features, like sandboxes or safety-tiered functions will further allow users to progressively grasp agent capabilities and hazards. Furthermore, interfaces must actively signal agent operations in sensitive contexts and mandate explicit user confirmation for high-consequence actions, especially if the agent's understanding of task severity is uncertain.

Technologically, reliable human-agent collaboration demands agents with accurate contextual comprehension to navigate dynamic GUIs and recover from errors. Critical research should also focus on equipping agents with an internalized `consequences calculus' that differentiates action priorities by discerning outcome severity, from trivial to critical. This calculus would then dynamically modulate agent autonomy, triggering stricter confirmation protocols or human intervention for high-risk operations. Finally, building collaborative trust requires inherently secure, privacy-preserving agent architectures with clear operational boundaries. These systems should actively address known vulnerabilities and progress towards certifiable safety.

From a socio-technical governance perspective, the CSCW community should design accountability frameworks that apportion agent-induced errors within collaborative workflows, accounting for shared human-agent intent and evolving concepts of distributed control. Proactive governance also should embed safety through innovative design and deployment strategies. For instance, product architectures that stratify functions by risk, or phased rollouts acting as `ecological sandboxes,' can facilitate gradual user adaptation and mitigate systemic risks.

\subsection{Limitations and Future Work}

This study's findings on UCs are subject to certain limitations. First, our investigation primarily centered on GUI agents within web browsing tasks. We did not explicitly differentiate UCs between desktop and mobile platforms, as the early developmental stage of GUI agent applications, with consequently limited mature and diverse cross-platform use cases, restricted our ability to observe and analyze potential platform-specific distinctions. Our specific focus, while enabling an in-depth analysis, means that the full spectrum of unintended behaviors potentially occurring with agents in other specialized domains or those utilizing non-textual interfaces, such as voice or gesture, was not comprehensively explored. Second, the research adopted an exploratory, qualitative methodology. While effective for identifying diverse novel issues and user experiences, this approach did not involve direct observations and controls of GUI agent use in lab settings, a methodological choice aligned with capturing broad real-world phenomena with agent technologies. It was also not designed to quantify prevalence or establish causal links, leaving an open area for quantitative complementary validation.
% Although this approach proved effective for identifying a diverse range of novel issues and user experiences, it was not designed to quantify the prevalence of these consequences or to establish definitive causal links under strictly controlled experimental conditions, leaving an open area for complementary quantitative validation.

\section{Conclusions}

The proliferation of LLM-based GUI agents in web browsing environments introduces significant UCs, challenging user trust, operational reliability and societal acceptance. Through a mixed-methods approach, we systematically characterized these phenomena (RQ1), revealing pervasive issues including privacy violations during input, unanticipated agent actions, the amplification of misinformation in outputs, and overall task failures. These operational breakdowns translate into critical multi-level influences (RQ2), leading to tangible harms such as security and privacy risks, negative user experiences including frustration, and an erosion of trust that hinders wider adoption. While users currently employ various mitigation strategies like manual corrections and configuring isolated environments (RQ3), these often serve as imperfect workarounds, underscoring profound implications to rethink human-GUI agent collaboration.% . Ultimately, fostering responsible and effective human-agent collaboration necessitates both technical advancements and a fundamental rethinking of interaction paradigms to balance automation with crucial human oversight.

% \begin{acks}
% \end{acks}

\bibliographystyle{ACM-Reference-Format}
\bibliography{sample}

\appendix

\section{Ethical Considerations}\label{appendix:ethical}

We acknowledge the potential ethical concerns of our research. Our study design and ethical considerations adhered to the principles outlined in the Menlo Report~\cite{bailey2012menlo} and the Belmont Report~\cite{beauchamp2008belmont}, ensuring a focus on responsibility, beneficence, and justice. The Institutional Review Board (IRB) of our institution approved all experimental procedures. Our analysis of Reddit data strictly adhered to the platform's terms of service, and we did not disclose any personal data or user profiles associated with the Reddit content. Before starting the experiment, we provided participants with a consent form and informed them of their right to request selective anonymization or to refrain from disclosing experiences if they felt any aspect was unfair or if they were uncomfortable sharing. Participants were explicitly informed of their right to withdraw from the experiment at any time without penalty. We stored all original experimental data in encrypted format on a secure local server at our institution.

\section{Details of Social Media Analysis}\label{appendix:social}

Our social media analysis covered posts from different subreddits, with 44 from r/OpenAI, 42 from r/ClaudeAI, 21 from r/singularity, 18 from r/LocalLLaMA, 14 from r/AI\_Agents, 8 from r/ArtificialIntelligence, 6 from r/ChatGPT, 5 from r/ChatGPTCoding, 5 from r/Anthropic, and others from other subreddits.

\section{Semi-structured Interview Protocol}
\label{appendix: interview protocol}

This protocol is designed for semi-structured interviews, allowing for both predefined questions and follow-up questions based on participant responses. Examples are only used when participants want further clarification on the questions.

\subsection{Section 1: Basic Information}

1. Could you describe your first experience using a GUI Agent, or your most memorable experience with a GUI Agent? What prompted you to start using GUI Agents?

2. In what situations do you typically use GUI Agents? 

3. Could you describe a recent task you completed using a GUI Agent in detail? 

4. Have you ever used GUI Agents for browser automation tasks? If so, could you describe one of your most recent experiences?

5. What different GUI Agents have you used? How would you compare them in terms of the following factors: functionality, usability, privacy and security concerns, and efficiency?

\subsection{Section 2: Unexpected Behaviors and Impacts}

6. Have you ever encountered any unexpected behavior from a GUI Agent? If yes, could you describe the most memorable instance? (e.g., did the Agent perform tasks you did not request, or execute tasks in an unexpected way?)

7. Have you ever experienced a situation where a GUI Agent was unable to complete a task? If yes, could you describe the most memorable instance in detail?

8. Have you ever received incorrect or misleading information from a GUI Agent? If yes, could you provide some specific examples?

9. Have you encountered any privacy or security issues related to GUI Agents? (e.g., did an Agent access your personal information without authorization, or did its behavior result in data leakage?)

10. Which of these unexpected behaviors had the most significant impact on you? What were the consequences? (e.g., time loss, financial loss, privacy breaches, security risks, decreased user experience)

11. Did you later discover the reason for the unexpected behavior of the GUI Agent? Or do you still not know the cause? If you discovered it, what was the reason, and how did you find out? Was it due to configuration errors, limited model capabilities, insufficient context understanding, or system complexity?

\subsection{Section 3: User Coping Strategies}

12. What measures have you taken to prevent or respond to these unexpected behaviors when using GUI Agents? (e.g., limiting Agent permissions, manually checking Agent operations, using virtual machines or containers to run Agents)

13. Do you think your coping strategies were effective? If yes, to what extent? 

14. What problems do you still encounter after the strategies you adopted? 

15. Do you know why these strategies were ineffective or why problems persisted? If not, do you have any guesses?

\subsection{Section 4: User Expectations for Developers}

16. What functions or features do you think GUI Agents should offer to reduce unexpected behaviors? 

17. How could they help you better control the Agent's actions and prevent unexpected behaviors?

\subsection{Section 5: Future Outlook}

18. What do you think is the future trend of GUI Agents? 

19. How do you think GUI Agents will change our work and lifestyle?

20. What are your expectations for the future development of GUI Agents? What new functions would you like GUI Agents to achieve?

21. Is there anything else you would like to add?

\end{document}